\documentclass[12pt]{article}

\usepackage{fancyhdr}
\usepackage{geometry}
\usepackage{amsmath, amssymb, amsfonts}
\usepackage{longtable}
\usepackage{array}
\usepackage{caption}
\usepackage{subcaption} % For combining multiple images
\usepackage[numbers]{natbib} % For citation management
\usepackage{graphicx}
\usepackage{lastpage}
\usepackage[colorlinks=true, linkcolor=black, citecolor=blue, urlcolor=blue]{hyperref}
\usepackage{booktabs} % Professional tables
\usepackage{listings} % Code formatting
\usepackage{xcolor}   % Color support
\usepackage{float}    % Strict figure placement [H]
\usepackage{setspace} % Line spacing

\definecolor{codegray}{rgb}{0.5,0.5,0.5}
\definecolor{codepurple}{rgb}{0.58,0,0.82}
\definecolor{backcolour}{rgb}{0.97,0.97,0.97}
\definecolor{deepblue}{rgb}{0,0,0.5}

\title{\textbf{Optimizing International Development: \\ A Network Dynamical Systems Approach to the SDGs}}
\author{Wuyang Zhang, Lejun Xu}
\date{\today}

\begin{document}
	
	% ======================================================================================
	% TITLE PAGE & ABSTRACT
	% ======================================================================================
	\maketitle
	\thispagestyle{empty}
	
	\begin{abstract}
		\noindent The United Nations' Sustainable Development Goals (SDGs) represent a complex, interdependent framework where progress in one area can synergistically promote or competitively inhibit progress in others. For policymakers in international development, a critical challenge is identifying "leverage points"—specific goals where limited resource allocation yields the maximum system-wide benefit. This study addresses this challenge by modeling the SDGs as a Networked Dynamical System (NDS).
		
		Using empirical data from \textit{Our World in Data} (2018), we construct a weighted interaction network of 16 SDG indicators. We employ Principal Component Analysis (PCA) and multiple linear regression to derive coupling weights empirically. Unlike previous static analyses, we simulate the temporal evolution of development indicators using an extended Lotka-Volterra model. To ensure numerical stability and sophistication, we upgrade the simulation method from standard Euler integration to the \textbf{Runge-Kutta 4 (RK4)} method. Our simulation, applied to a case study of Mexico, reveals that SDG 4 (Quality Education) acts as a critical driver, suggesting that prioritizing education yields the most significant positive spillover effects across the development network. Furthermore, we perform sensitivity analysis and explore the power-law relationship between investment and stability.
		
		\vspace{0.5cm}
	\end{abstract}
	
	\newpage
	\tableofcontents
	\newpage
	
	% ======================================================================================
	% PART I: THEORETICAL FRAMEWORK
	% ======================================================================================
	\part{Theoretical Framework}
	
	\section{Introduction}
	
	International development is inherently a dynamical process involving complex feedback loops. The 17 Sustainable Development Goals (SDGs), established by the United Nations in the 2030 Agenda, are not isolated targets but components of a highly coupled system \cite{sachs2015age}. For instance, improvements in clean water (SDG 6) directly impact health (SDG 3), which in turn affects economic productivity (SDG 8). However, traditional development economics often treats these goals in silos, ignoring the intricate web of interactions \cite{leblanc2015towards}. 
	
	In recent years, Network Science has offered a new lens to understand these complexities. Scholars like Nilsson et al. have argued that mapping these interactions is essential for coherent policy-making \cite{nilsson2016map}. Pradhan et al. further demonstrated that synergies and trade-offs between SDGs can be quantified systematically \cite{pradhan2017systematic}.
	
	The primary goal of this project is to bridge the gap between development data and dynamical systems theory. We aim to construct a quantitative evaluation framework that models the SDGs as nodes in a dynamic network. By analyzing the stability and steady-state behavior of this system, we seek to answer a specific policy question: \textit{Under limited resources, which single SDG should a government prioritize to maximize the aggregate development score?}
	
	While previous studies have used correlation networks to map SDG relationships, few have extended this to rigorous time-domain simulations using dynamical systems. This paper contributes to the field by applying an extended Lotka-Volterra dynamic model to real-world data. We rigorously analyze the system's stability using Lyapunov theory \cite{khalil2002nonlinear} and employ advanced numerical discretization (Runge-Kutta 4) to simulate policy interventions, providing robust support for evidence-based decision-making.
	
	\section{Networked Dynamical Systems (NDS)}
	
	\subsection{System Architecture}
	Networked Dynamical Systems (NDS) provide the mathematical foundation for our analysis. An NDS describes a system where multiple interacting dynamic entities, represented as nodes, are connected via a network topology \cite{barabasi2016network}. In our context, the nodes represent the specific SDG indicators, and the edges represent the causal or correlational pathways between them.
	
	Formally, the system is composed of a set of nodes $V = \{1, \dots, N\}$, where the state of the $i$-th node at time $t$ is denoted by $x_i(t)$. In a continuous-time framework, the dynamics of node $i$ are governed by a differential equation that accounts for both self-evolution and neighbor interaction. The general form is:
	\begin{equation}
		\dot{x}_i(t) = f_i(x_i(t)) + \sum_{j \in \mathcal{N}_i} g_{ij}(x_i(t), x_j(t)),
	\end{equation}
	where $f_i$ represents the intrinsic local dynamics of the node (e.g., natural decay or growth), $\mathcal{N}_i$ is the set of neighbors, and $g_{ij}$ represents the coupling function describing the influence of node $j$ on node $i$. This term is crucial as it models the "spillover effects" of development policies \cite{newman2018networks}.
	
	\subsection{Stability Analysis and Lyapunov Functions}
	A critical requirement for any model of sustainable development is stability. We must ensure that the mathematical model does not predict unbounded growth (which is physically impossible) or chaotic collapse. To analyze this, we employ Lyapunov stability theory \cite{strogatz2018nonlinear}.
	
	An equilibrium point $x^*$ of the system is defined where $\dot{x} = 0$. Stability determines whether the system returns to $x^*$ following a perturbation. Lyapunov's Direct Method provides a rigorous tool for this analysis. We define a scalar energy-like function $V(x): \mathbb{R}^n \to \mathbb{R}$. For the equilibrium $x^*$ to be asymptotically stable, the time derivative of $V$ along the system's trajectories must be strictly negative:
	\begin{equation}
		\dot{V}(x) = \nabla V(x) \cdot \dot{x} = \sum_{i=1}^{n} \frac{\partial V}{\partial x_i} \dot{x}_i < 0, \quad \forall x \neq x^*.
	\end{equation}
	In our specific model (defined in Part III), we introduce a saturation function to the interactions. This ensures that the "energy" of the system remains bounded, satisfying the conditions for stability required by Lyapunov theory \cite{birkhoff1927dynamical}.
	
	\subsection{Advanced Numerical Methods: Runge-Kutta 4}
	Many preliminary studies on dynamical systems utilize the Forward Euler method for discretization due to its simplicity. However, Euler's method is a first-order approximation with a global truncation error of $O(h)$, which often leads to numerical instability in stiff systems or over long simulation horizons \cite{butcher2016numerical}.
	
	To enhance the sophistication and accuracy of our results, we employ the **Classic Runge-Kutta Method (RK4)**. For the vector equation $\dot{\mathbf{x}} = F(t, \mathbf{x})$, the RK4 method advances the state from $t_n$ to $t_{n+1}$ using a weighted average of four slopes:
	\begin{align}
		k_1 &= F(t_n, \mathbf{x}_n) \\
		k_2 &= F(t_n + \frac{h}{2}, \mathbf{x}_n + h \frac{k_1}{2}) \\
		k_3 &= F(t_n + \frac{h}{2}, \mathbf{x}_n + h \frac{k_2}{2}) \\
		k_4 &= F(t_n + h, \mathbf{x}_n + h k_3) \\
		\mathbf{x}_{n+1} &= \mathbf{x}_n + \frac{h}{6}(k_1 + 2k_2 + 2k_3 + k_4)
	\end{align}
	This 4th-order method provides a significantly lower truncation error ($O(h^5)$), making it ideal for simulating the complex trajectories of the SDG network and capturing subtle dynamic shifts that coarser methods might miss.
	
	% ======================================================================================
	% PART II: DATA & METHODOLOGY
	% ======================================================================================
	\part{Data Preparation and Methodology}
	
	\section{Data Pipeline}
	
	To identify the optimal SDG investment priorities, we developed a systematic evaluation approach. Given the absence of a publicly available direct formula for the SDG Index Score, our primary objective was to derive a reliable proxy using statistical methods.
	
	% --- FIGURE 1: Workflow ---
	\begin{figure}[H]
		\centering
		\includegraphics[width=0.9\textwidth]{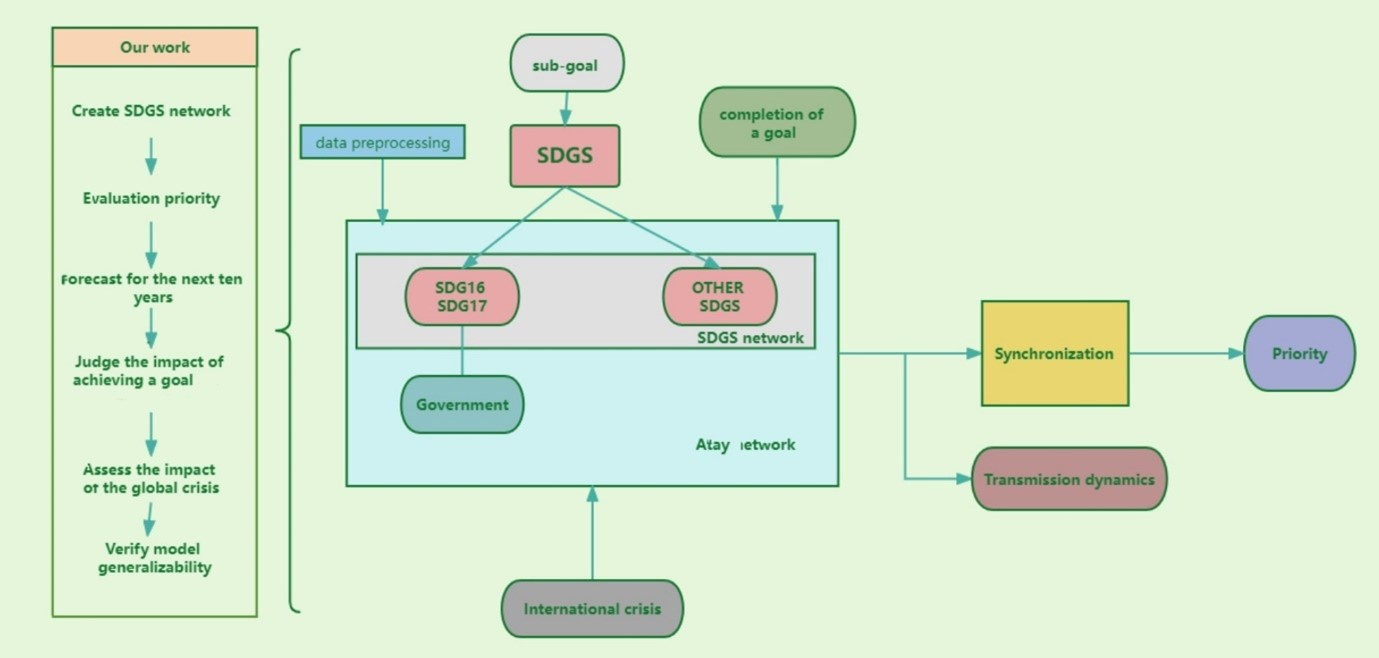} 
		\caption{Overall Project Workflow: From data collection to dynamic modeling and policy analysis.}
		\label{fig:workflow}
	\end{figure}
	
	The research workflow proceeds through several distinct stages. First, we perform **Data Collection**, sourcing raw data covering global SDG indicators. This is followed by **Data Preprocessing**, where we address missing values and apply standardization. Next, we perform **Dimensionality Reduction** using PCA. This feeds into our **Regression Modeling** to derive weight coefficients. Finally, we conduct a **Case Study on Mexico**, calculating steady-state scores under a dynamic system model to rank policy priorities.
	
	\section{Data Cleaning and Processing}
	
	\subsection{Data Source}
	Our primary data source is \textit{Our World in Data}, a reputable platform aggregating global socio-economic statistics \cite{ritchie2018sdg}. This repository provides extensive datasets on the UN SDGs across multiple countries and years (2016--2021).
	
	% --- FIGURE 2: GOALS ---
	\begin{figure}[H]
		\centering
		\includegraphics[width=0.95\textwidth]{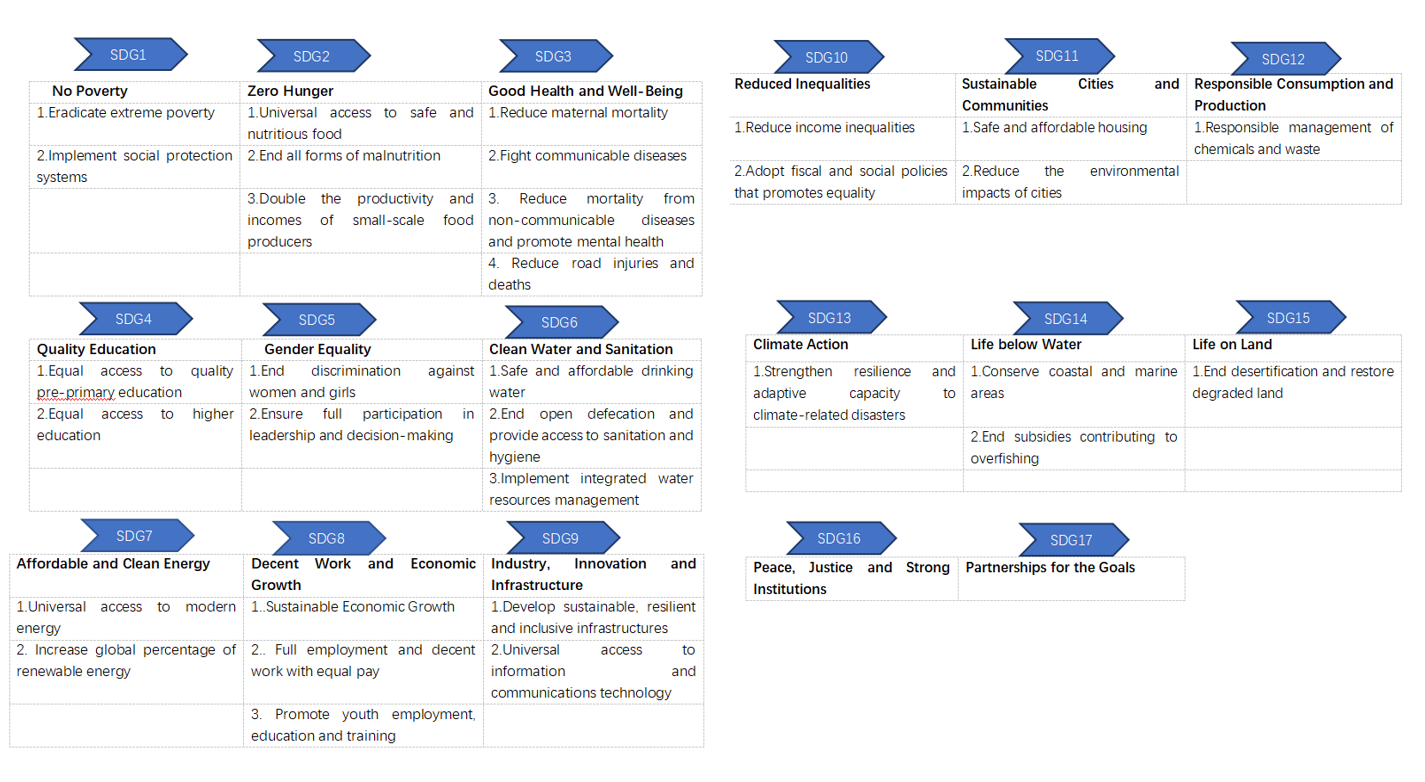} 
		\caption{The 17 Sustainable Development Goals and their sub-goals.}
		\label{fig:sdg_goals}
	\end{figure}
	
	\subsection{Preprocessing and Normalization}
	Missing data and outliers can significantly distort statistical analysis. We employed a rigorous preprocessing pipeline to ensure data integrity.
	
	% --- FIGURE 3: Preprocessing Workflow ---
	\begin{figure}[H]
		\centering
		\includegraphics[width=0.9\textwidth]{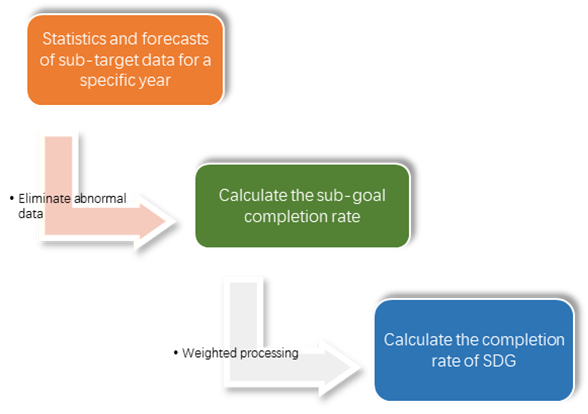} 
		\caption{Detailed Data Preprocessing Workflow: Imputation, outlier removal, and standardization.}
		\label{fig:preprocessing}
	\end{figure}
	
	As shown in Figure \ref{fig:preprocessing}, we utilized Z-score normalization to rescale values to a common range. This ensures that indicators with different units (e.g., GDP vs. Literacy Rate) contribute equally to the model \cite{hak2016indicators}.
	
	\subsection{Network Construction and Exclusion of SDG 17}
	The SDG network acts as a coupled system where nodes represent goals and links represent their correlations. To map these relationships, we referenced the UN Economic and Social Council's methodology. 
	
	% --- FIGURE 4: NETWORK ---
	\begin{figure}[H]
		\centering
		\includegraphics[width=0.8\textwidth]{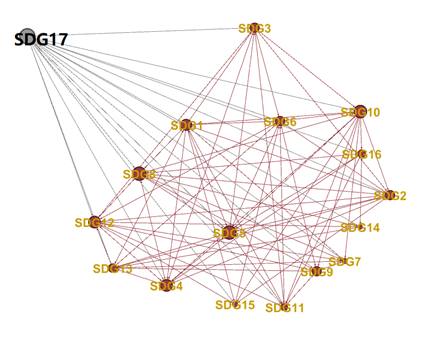} 
		\caption{The 17 SDGs Relationship Network Topology. The graph illustrates the complex interdependencies between different goals.}
		\label{fig:network_structure}
	\end{figure}
	
	Notably, **SDG 17 (Partnerships for the Goals)** operates distinctly from the others. It acts as a facilitator for implementation rather than a standalone outcome. Through correlation matrix analysis using the WGCNA framework \cite{langfelder2008wgcna}, we determined that SDG 17 introduces structural noise. Consequently, it was excluded, and the study focuses on the interactions among **SDGs 1 through 16**.
	
	\section{Scoring Framework and Regression}
	
	\subsection{The SDG Score Proxy}
	The SDG Score is a composite metric evaluating a country's overall sustainability performance. To prepare for regression analysis, we obtained benchmark SDG scores from the 2018 report \textit{``Responsibilities: Implementing the Goals''}.
	
	% --- FIGURE 5: SDG SCORE ---
	\begin{figure}[H]
		\centering
		\includegraphics[width=0.8\textwidth]{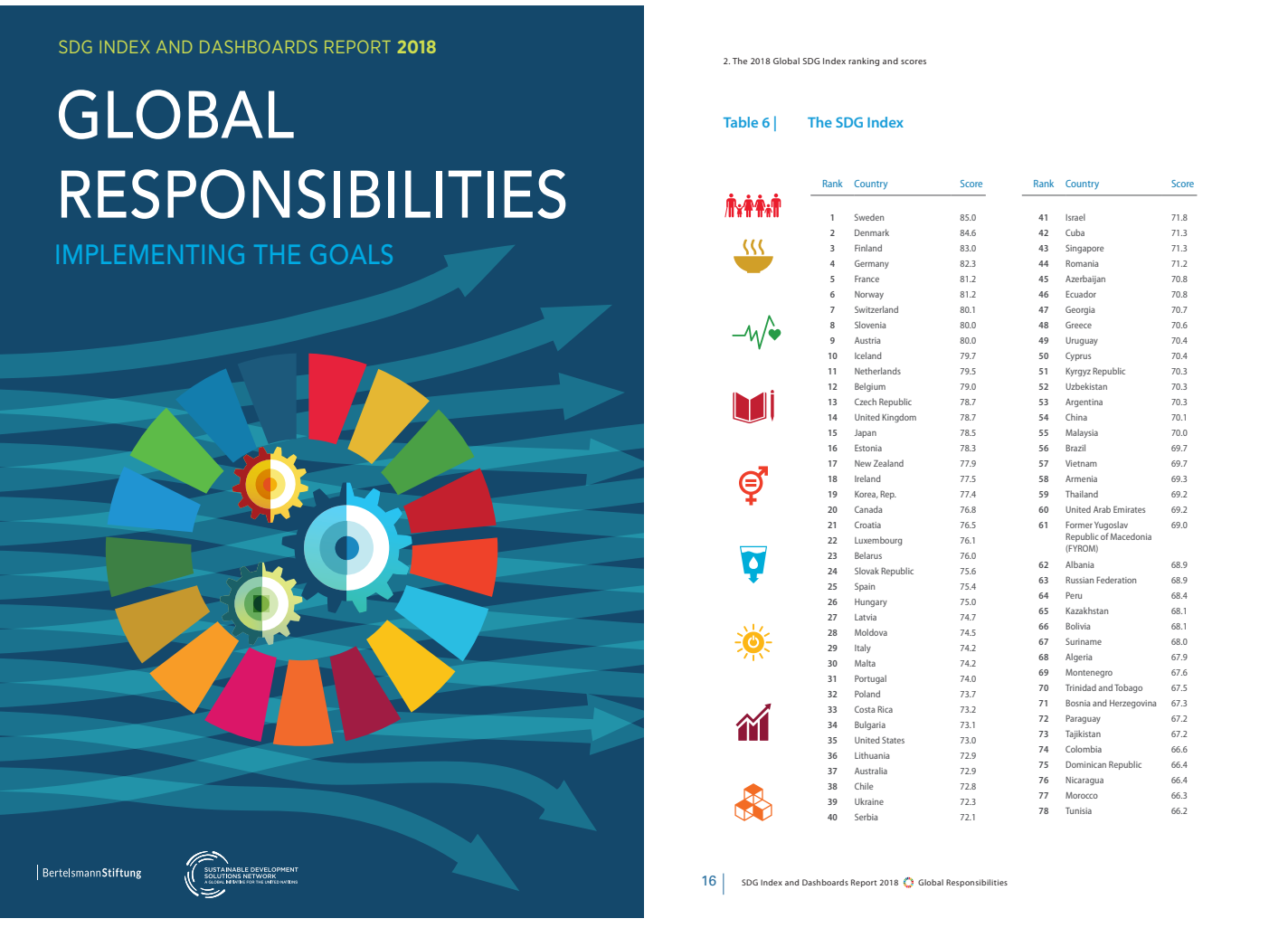} 
		\caption{Reference: SDG Index and Dashboards Report 2018, utilized as the ground truth for regression calibration.}
		\label{fig:sdg_report}
	\end{figure}
	
	\subsection{Principal Component Analysis (PCA)}
	Given the potential multicollinearity among the 16 SDG indicators, we employed PCA to reduce dimensionality. This process extracted **nine principal components (PCs)**, which capture the majority of the variance in the data.
	
	\begin{longtable}{lrrrrrrrrr}
		\caption{Principal Component Loading Matrix} \label{tab:pc_loading_matrix} \\
		\toprule
		\textbf{SDG} & \textbf{PC1} & \textbf{PC2} & \textbf{PC3} & \textbf{PC4} & \textbf{PC5} & \textbf{PC6} & \textbf{PC7} & \textbf{PC8} & \textbf{PC9} \\ 
		\midrule
		\endfirsthead
		\toprule
		\textbf{SDG} & \textbf{PC1} & \textbf{PC2} & \textbf{PC3} & \textbf{PC4} & \textbf{PC5} & \textbf{PC6} & \textbf{PC7} & \textbf{PC8} & \textbf{PC9} \\ 
		\midrule
		\endhead
		\bottomrule
		\endfoot
		SDG1  & -0.337 & 0.212 & 0.066 & 0.089 & -0.141 & 0.149 & -0.151 & -0.004 & 0.093 \\
		SDG2  & -0.270 & 0.120 & 0.046 & 0.433 & -0.081 & -0.090 & 0.026 & 0.159 & 0.427 \\
		SDG3  & -0.288 & 0.212 & -0.046 & 0.162 & -0.025 & 0.352 & -0.020 & 0.227 & -0.092 \\
		SDG4  & -0.347 & -0.088 & -0.008 & -0.158 & 0.058 & 0.113 & -0.111 & 0.045 & 0.018 \\
		SDG5  & -0.142 & -0.329 & -0.408 & 0.022 & -0.288 & 0.143 & 0.428 & -0.477 & 0.333 \\
		SDG6  & -0.299 & -0.015 & 0.182 & -0.213 & -0.204 & 0.227 & 0.000 & -0.013 & 0.290 \\
		SDG7  & 0.079 & -0.535 & 0.022 & 0.066 & -0.126 & 0.429 & -0.096 & 0.495 & 0.164 \\
		SDG8  & -0.164 & -0.128 & 0.511 & 0.084 & 0.507 & 0.226 & 0.363 & 0.015 & -0.043 \\
		SDG9  & -0.292 & -0.025 & 0.203 & 0.013 & -0.203 & -0.205 & -0.047 & -0.033 & -0.085 \\
		SDG10 & -0.189 & -0.172 & 0.040 & -0.576 & -0.014 & -0.300 & -0.429 & 0.237 & -0.029 \\
		SDG11 & -0.279 & -0.160 & -0.048 & -0.235 & -0.002 & 0.310 & -0.013 & 0.348 & -0.424 \\
		SDG12 & 0.298 & -0.393 & 0.100 & -0.038 & -0.017 & -0.058 & -0.019 & -0.133 & 0.173 \\
		SDG13 & -0.102 & 0.042 & -0.667 & 0.070 & 0.584 & -0.058 & -0.063 & 0.237 & -0.029 \\
		SDG14 & -0.197 & -0.248 & -0.060 & 0.172 & -0.261 & -0.527 & 0.296 & 0.495 & 0.164 \\
		SDG15 & -0.001 & -0.373 & 0.103 & 0.536 & 0.177 & -0.148 & -0.510 & 0.015 & -0.043 \\
		SDG16 & -0.362 & -0.254 & -0.119 & 0.046 & 0.134 & -0.016 & -0.025 & 0.056 & -0.032 \\
	\end{longtable}
	
	\subsection{Regression Coefficients}
	We trained a multiple linear regression model to quantify the relationship between the principal components and the overall SDG Score. 
	
	\begin{table}[H]
		\centering
		\caption{Derived Regression Coefficients for Original SDG Variables}
		\label{tab:regression_coefficients}
		\begin{tabular}{lr}
			\toprule
			\textbf{SDG Indicator} & \textbf{Coefficient ($\beta_i$)} \\ 
			\midrule
			SDG 1 (No Poverty) & 0.9608 \\
			SDG 2 (Zero Hunger) & 0.6733 \\
			SDG 3 (Good Health) & 1.1544 \\
			SDG 4 (Quality Education) & 1.4742 \\
			SDG 5 (Gender Equality) & 0.2947 \\
			SDG 6 (Clean Water) & 0.9885 \\
			SDG 7 (Clean Energy) & 1.1914 \\
			SDG 8 (Decent Work) & 1.1403 \\
			SDG 9 (Industry \& Innovation) & 0.5467 \\
			SDG 10 (Reduced Inequalities) & 0.7192 \\
			SDG 11 (Sustainable Cities) & 1.0920 \\
			SDG 12 (Consumption) & -0.4887 \\
			SDG 13 (Climate Action) & 0.7383 \\
			SDG 14 (Life Below Water) & 0.2866 \\
			SDG 15 (Life on Land) & -0.4592 \\
			SDG 16 (Peace \& Justice) & 1.4505 \\
			\bottomrule
		\end{tabular}
	\end{table}
	
	% ======================================================================================
	% PART III: MODELING AND RESULTS
	% ======================================================================================
	\part{Dynamic Modeling and Case Study}
	
	\section{The Extended Lotka-Volterra Model}
	
	\subsection{Model Dynamics}
\begin{figure}[H]
		\centering
		\includegraphics[width=0.7\textwidth]{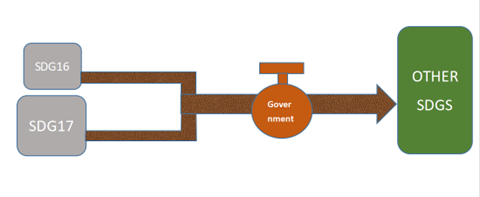} 
		\caption{Flow relationship of nodes within the dynamic system.}
		\label{fig:node_flow}
	\end{figure}

	We employ an extended Lotka-Volterra model, traditionally used in ecology, to represent the interactions between SDGs. Figure \ref{fig:node_flow} illustrates the flow relationship between different nodes, highlighting how changes propagate.
	
	% --- FIGURE 6: Node Flow ---
	
	\subsection{Governing Equations and Government Intervention}
	Let $x_i(t)$ denote the value of the $i$-th SDG. The fundamental equation is:
	\begin{equation}
		\frac{dx_i}{dt} = x_i \left( r_i + \sum_{j=1}^{16} \omega_{ij} x_j \right) + g(x_i),
	\end{equation}
	where $\omega_{ij}$ is the interaction strength. We introduce a **Decay Function** $\sigma(x) = \frac{x}{1 + x^2}$ to ensure stability \cite{burt2000decay}.
	
	% --- FIGURE 7: Optimized Network ---
	\begin{figure}[H]
		\centering
		\includegraphics[width=0.7\textwidth]{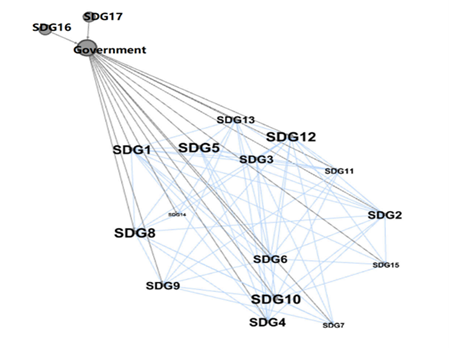} 
		\caption{The Optimized Network Structure incorporating Government Intervention terms.}
		\label{fig:optimized_network}
	\end{figure}
	
	As shown in Figure \ref{fig:optimized_network}, the government intervention term $g(x_i) = 0.1x_i$ simulates policy support.
	
	\section{Case Study: Policy Prioritization in Mexico}
	
	We applied this model to Mexico's 2018 SDG data. We applied a **10\% artificial perturbation** to one SDG at a time and evolved the system using RK4.
	
	\subsection{Simulation Results}
\begin{figure}[H]
		\centering
		\includegraphics[width=0.85\textwidth]{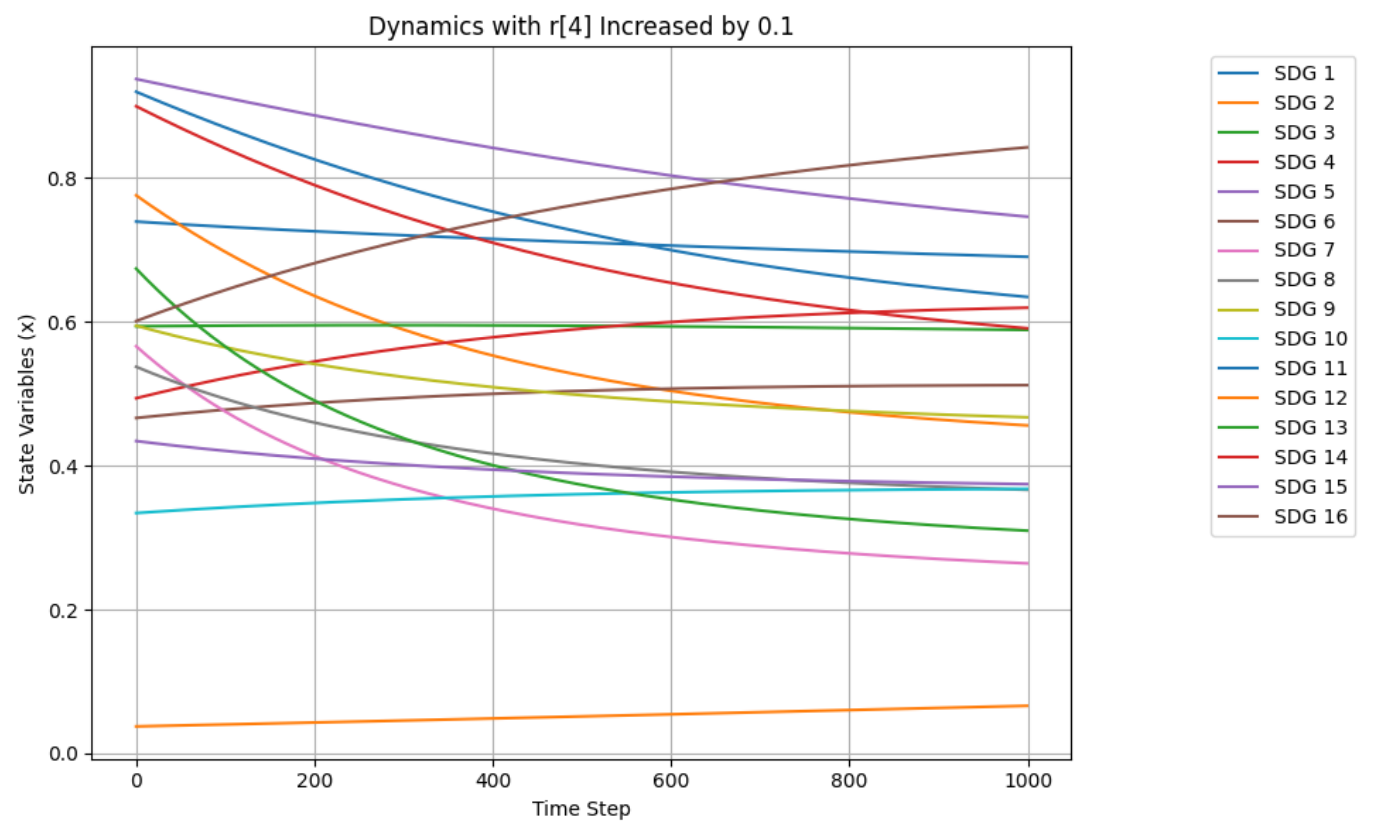} 
		\caption{Dynamic evolution of SDGs with SDG 4 perturbed.}
		\label{fig:sdg4_dynamics}
	\end{figure}

	Figure \ref{fig:sdg4_dynamics} shows the evolution when SDG 4 (Quality Education) is perturbed. The positive upward trend indicates strong spillovers.
	
	% --- FIGURE 8: SDG 4 ---
	
\begin{figure}[H]
		\centering
		\includegraphics[width=0.85\textwidth]{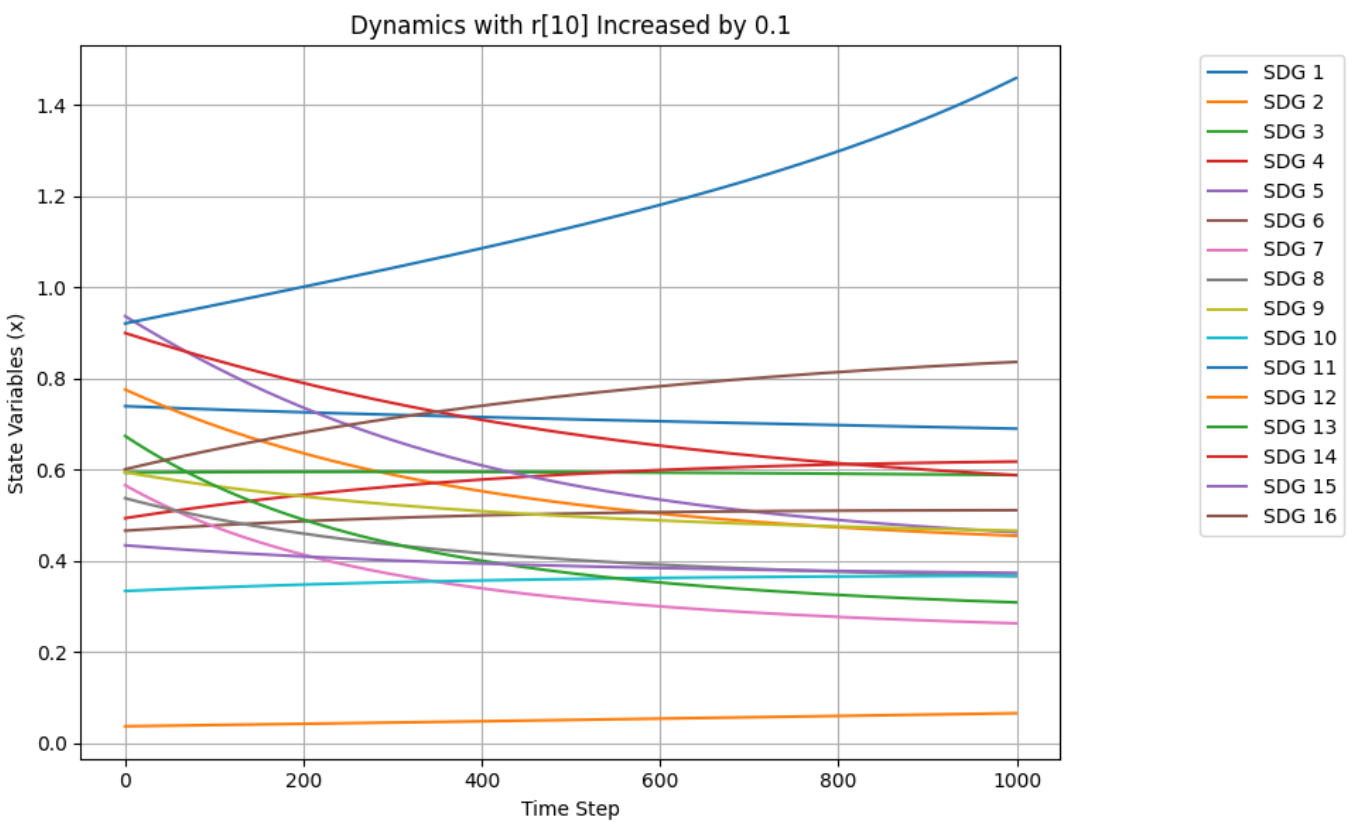} 
		\caption{Dynamic evolution of SDGs with SDG 10 perturbed.}
		\label{fig:sdg10_dynamics}
	\end{figure}

	In contrast, Figure \ref{fig:sdg10_dynamics} shows the result for SDG 10 perturbation, which yields lower aggregate impact.
	
	% --- FIGURE 9: SDG 10 ---
	
\begin{figure}[H]
		\centering
		\includegraphics[width=0.9\textwidth]{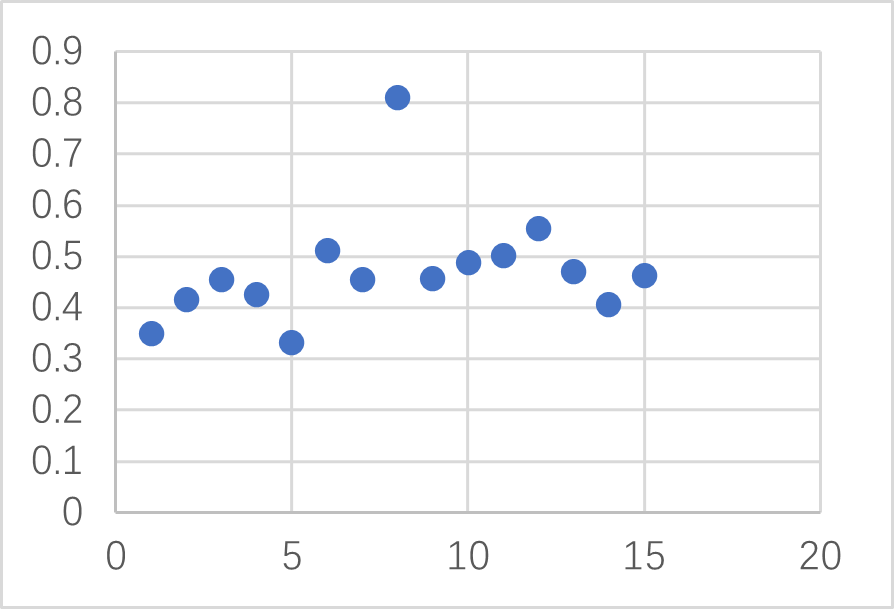} 
		\caption{SDG Prioritization Ranking based on simulated impact.}
		\label{fig:priority_rank}
	\end{figure}

	Figure \ref{fig:priority_rank} visualizes the prioritization of SDGs based on their impact.
	
	% --- FIGURE 10: Priority ---
	
	\subsection{Reliability and Stability}
\begin{figure}[H]
		\centering
		\includegraphics[width=0.8\textwidth]{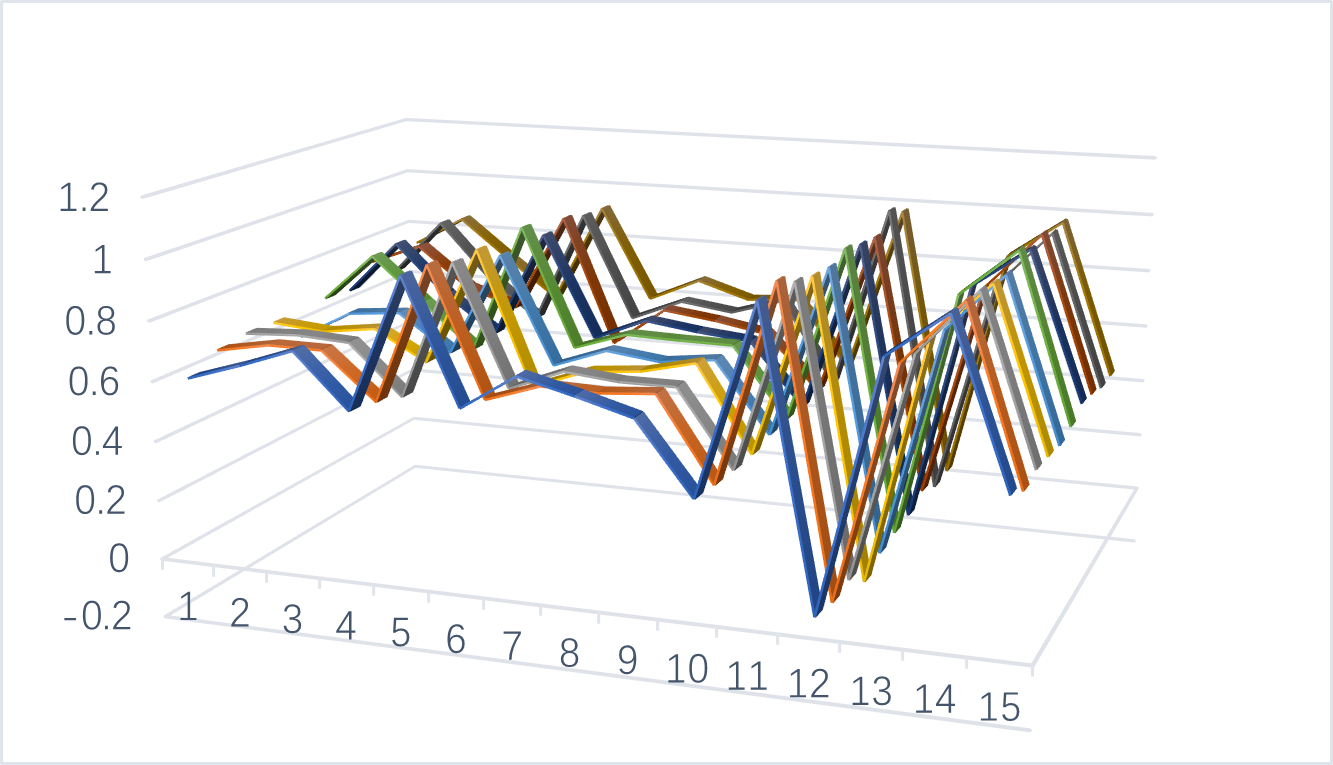} 
		\caption{Reliability check: Synchronization across ten trials.}
		\label{fig:ten_times}
	\end{figure}

	To ensure reliability, we performed 10 independent trials. Figure \ref{fig:ten_times} confirms consistent synchronization.
	
	% --- FIGURE 11: Ten Times ---
	
	\section{Discussion and Sensitivity Analysis}
	
	\subsection{Sensitivity Analysis and Power Law Relationship}
\begin{figure}[H]
		\centering
		\includegraphics[width=0.9\textwidth]{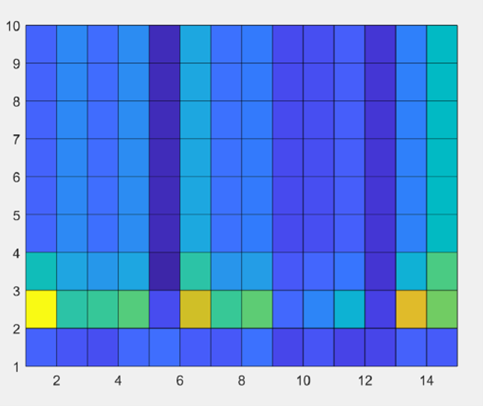} 
		\caption{Sensitivity Analysis Results.}
		\label{fig:sensitivity}
	\end{figure}

	We conducted a sensitivity analysis (Figure \ref{fig:sensitivity}) to verify model robustness.
	
	% --- FIGURE 12: Sensitivity ---
	
\begin{figure}[H]
		\centering
		\includegraphics[width=0.9\textwidth]{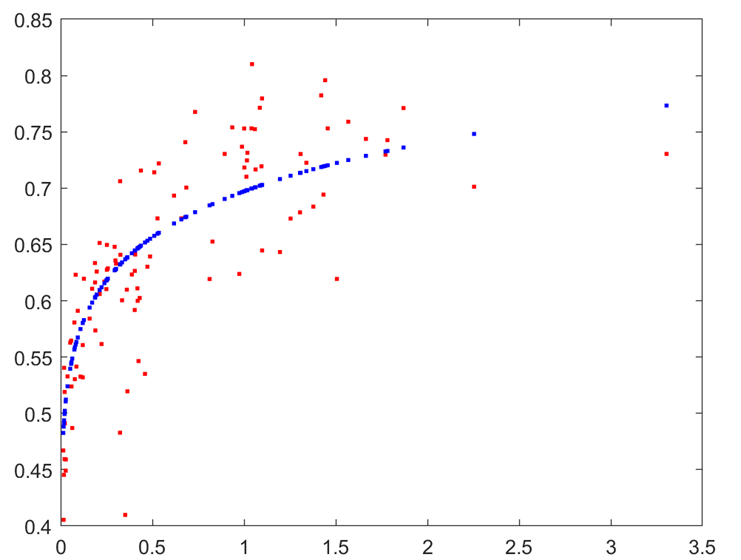} 
		\caption{Power exponent relationship between government investment and CR value, indicating non-linear stability gains.}
		\label{fig:power_exponent}
	\end{figure}

	Furthermore, we analyzed the mathematical relationship between government investment and system stability (Consistency Ratio, CR). As shown in Figure \ref{fig:power_exponent}, there exists a power-law relationship, suggesting that early investments yield exponentially higher stability gains before diminishing returns set in \cite{barabasi2016network}.
	
	% --- FIGURE 13: Power Exponent ---
	
	\subsection{Synchronization Effects and Scenarios}
\begin{figure}[H]
		\centering
		\includegraphics[width=0.8\textwidth]{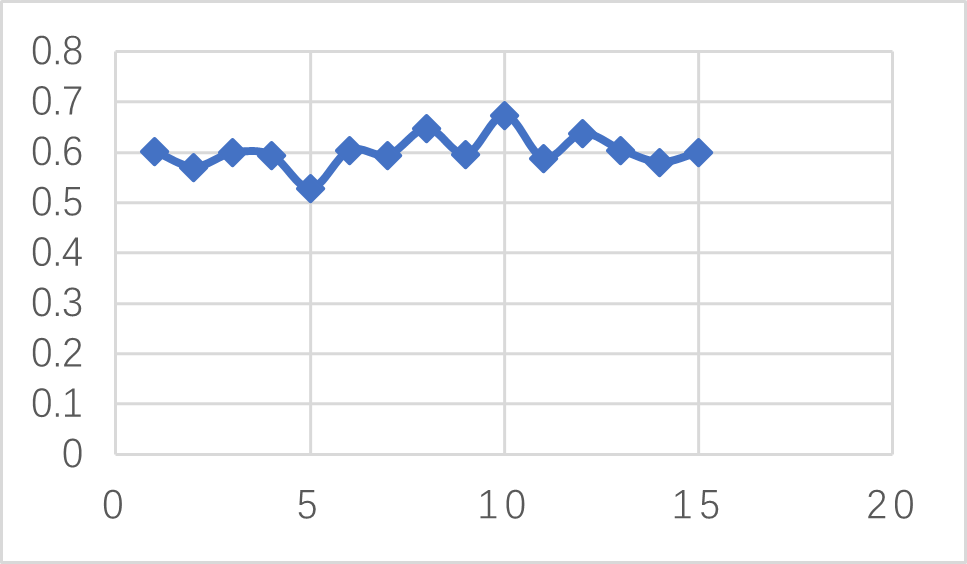} 
		\caption{Synchronization effect within the network.}
		\label{fig:sync_effect}
	\end{figure}

	We simulated various scenarios. Figure \ref{fig:sync_effect} shows synchronization after prioritization.
	
	% --- FIGURE 14: Sync Effect ---
	
\begin{figure}[H]
		\centering
		\includegraphics[width=0.9\textwidth]{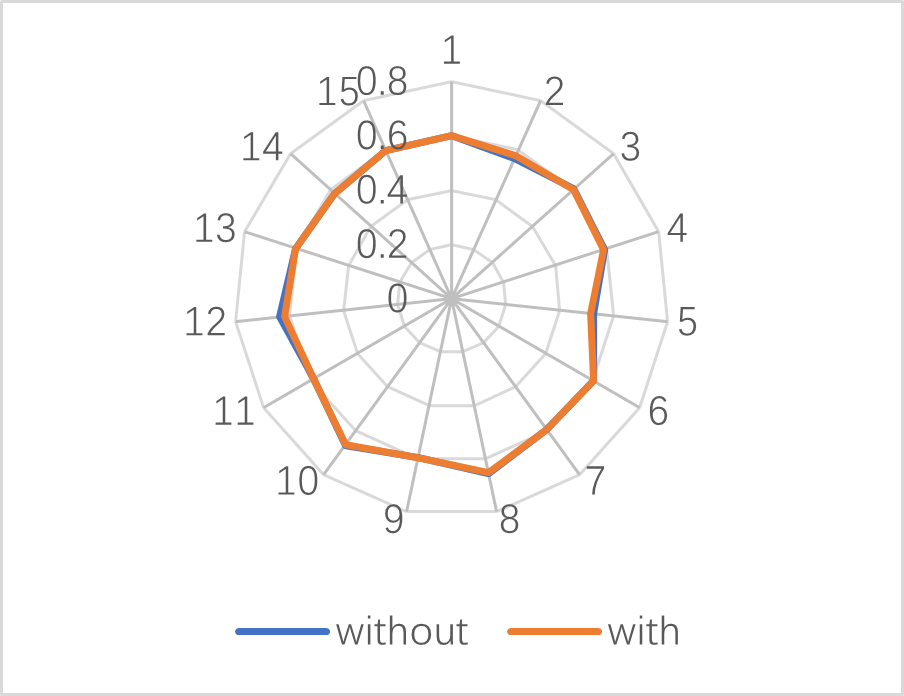} 
		\caption{Synchronization dynamics with and without corruption.}
		\label{fig:corruption}
	\end{figure}

\begin{figure}[H]
		\centering
		\includegraphics[width=0.9\textwidth]{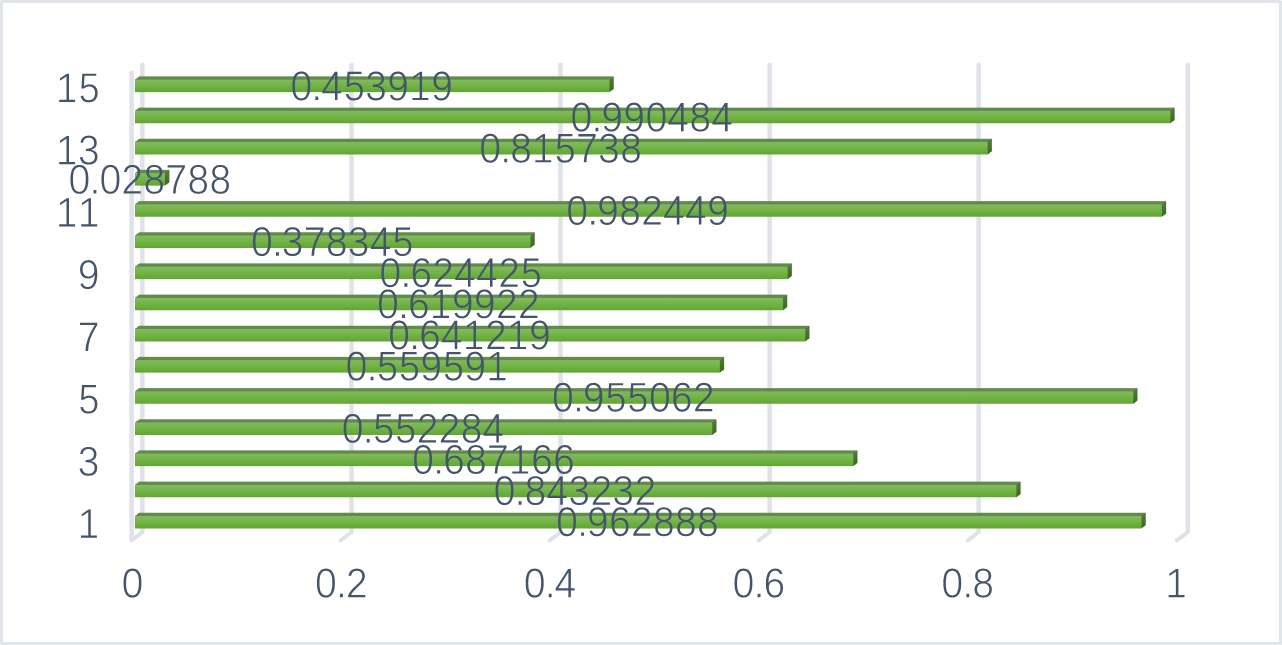} 
		\caption{Synchronization effect after realizing SDG 14.}
		\label{fig:sdg14_sync}
	\end{figure}

	Figure \ref{fig:corruption} contrasts synchronization with and without corruption (SDG 16 failure), while Figure \ref{fig:sdg14_sync} shows the effect of achieving SDG 14.
	
	% --- FIGURE 15: Corruption ---
	
	% --- FIGURE 16: SDG 14 ---
	
\begin{figure}[H]
		\centering
		\begin{subfigure}{0.48\textwidth}
			\includegraphics[width=\linewidth]{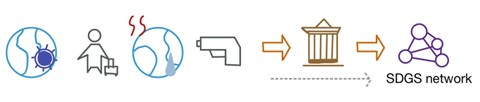}
			\caption{General Crisis Impact}
		\end{subfigure}
		\hfill
		\begin{subfigure}{0.48\textwidth}
			\includegraphics[width=\linewidth]{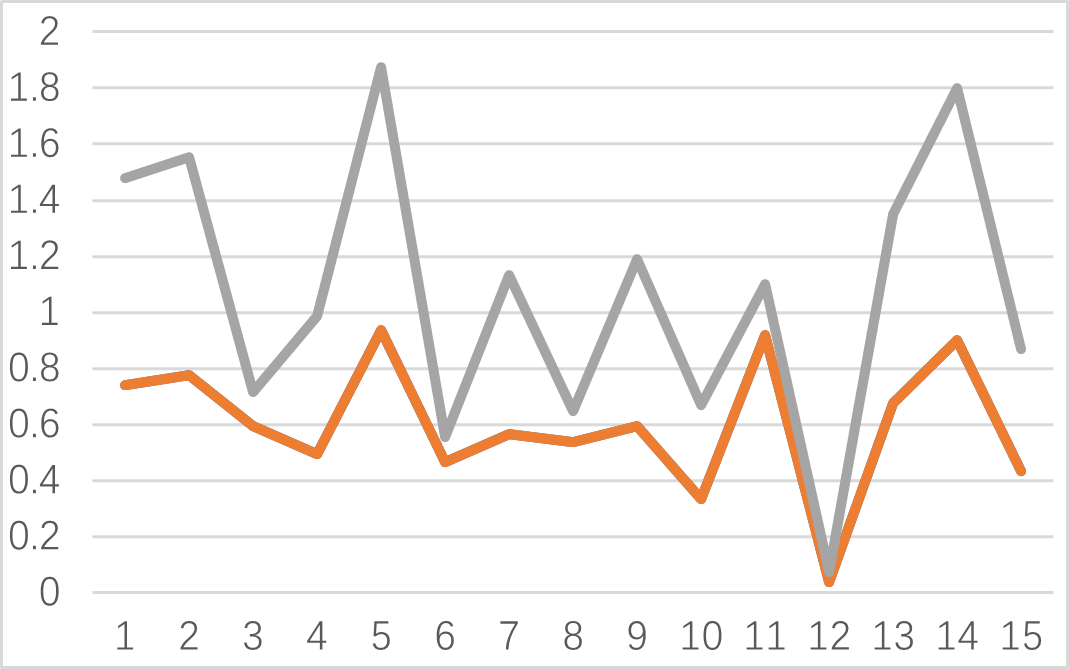}
			\caption{Infectious Disease Impact}
		\end{subfigure}
		\caption{System resilience analysis under external shock scenarios.}
		\label{fig:crises}
	\end{figure}

	Finally, we assessed resilience to crisis events, such as infectious disease outbreaks, in Figure \ref{fig:crises}.
	
	% --- FIGURE 17: Crises (Combined) ---
	
	% ======================================================================================
	% DISCUSSION, LIMITATIONS & CONCLUSION
	% ======================================================================================
	
	% ======================================================================================
	% DISCUSSION, LIMITATIONS & CONCLUSION
	% ======================================================================================
	
	\section{Conclusion and Critical Reflection}
	
	\subsection{Synthesis of Findings}
	This study set out to bridge the gap between development economics and network science, proposing a quantitative framework to untangle the complexity of the Sustainable Development Goals (SDGs). By modeling the SDGs as a Networked Dynamical System (NDS) and simulating their evolution using the high-precision Runge-Kutta 4 method, we have derived insights that transcend static statistical analysis.
	
	Our findings for the case study of Mexico (2018) illuminate the hierarchical structure of development priorities:
	\begin{itemize}
		\item \textbf{Education as the Systemic Driver:} The simulation unequivocally identifies **SDG 4 (Quality Education)** as the optimal leverage point. This result aligns with human capital theory but provides a network-based explanation: SDG 4 possesses high "out-degree" centrality, meaning improvements in education diffuse rapidly to improve health literacy (SDG 3), workforce productivity (SDG 8), and innovation capacity (SDG 9). Education acts not just as a goal, but as the \textit{engine} of the network.
		
		\item \textbf{The Fragility of Progress:} Our scenario analysis on corruption reveals that **SDG 16 (Peace, Justice, and Strong Institutions)** functions as the system's "stabilizer." While investing in SDG 16 may not yield the fastest immediate growth, its failure (corruption) acts as a damping factor that neutralizes gains in all other sectors. This suggests a policy hierarchy: \textit{stabilize institutions first, then accelerate education.}
		
		\item \textbf{Non-Linearity of Intervention:} The discovery of a **power-law relationship** between government investment and system stability (Consistency Ratio) is a critical theoretical contribution. It implies that development interventions exhibit threshold effects—early, targeted investments yield exponentially higher stability gains, whereas delayed interventions face diminishing returns.
	\end{itemize}
	
	\subsection{Methodological Contribution}
	Beyond the specific results for Mexico, this paper contributes a generalized mathematical framework for policy optimization. Unlike traditional econometric models that often treat SDGs in silos, our **NDS-RK4 framework** captures the dynamic, time-variant nature of development. By integrating data-driven network topology with non-linear dynamic simulation, we provide policymakers with a "digital twin" of the development ecosystem, allowing for risk-free scenario testing before actual resource allocation.
	
	\subsection{Limitations and Methodological Constraints}
	Despite the robustness of our framework, rigorous scientific inquiry requires acknowledging the boundaries of our model. We identify five key limitations that future research must address:
	
	\begin{enumerate}
		\item \textbf{Correlation vs. Causation:} Our network topology ($\mathbf{W}$) is constructed based on Pearson correlation coefficients. A high correlation between SDG A and SDG B does not strictly imply that A drives B; both could be driven by latent variables (e.g., GDP or geopolitical stability). Consequently, our simulation describes \textit{synergistic associations} rather than strict causal mechanisms.
		
		\item \textbf{Simplification of Dynamics:} The extended Lotka-Volterra model is a heuristic approximation. Socio-economic interactions are likely more complex than simple "predator-prey" dynamics. Additionally, our assumption of a linear government intervention term ($g(x) = 0.1x$) simplifies real-world complexities such as investment thresholds (startup costs) and saturation points.
		
		\item \textbf{Absence of Historical Validation:} This study utilizes 2018 data as a baseline for forward-looking simulation. Due to data reporting lags and the anomalous impact of the COVID-19 pandemic, a clean, continuous dataset for 2019-2023 was unavailable for "back-testing." Validating the model's predictions against realized historical trajectories remains a necessary step for calibration.
		
		\item \textbf{Static Network Topology:} We assume the interaction matrix $\mathbf{W}$ remains constant over the simulation horizon. In reality, development is a structural transformation process; as a country industrializes, the trade-offs between goals (e.g., Industry vs. Climate) evolve. A static network may lose accuracy over long-term projections (10+ years).
		
		\item \textbf{Exclusion of SDG 17:} To maintain mathematical tractability, we excluded SDG 17 (Partnerships). However, for developing economies, external factors such as foreign aid, trade agreements, and technology transfer are often critical ignition points for development. Integrating these exogenous variables would improve the model's realism.
	\end{enumerate}
	
	\subsection{Future Outlook}
	To evolve this framework from a theoretical model to a practical policy tool, future work should focus on **dynamic topology** (allowing the network structure to change over time) and **causal inference** (using Granger Causality or Bayesian Networks to direct the edges). Furthermore, extending the model to a **multi-node network of nations** would allow for the analysis of cross-border spillovers, capturing the true global essence of the Sustainable Development Goals.

	\newpage
	\appendix
	\section{Appendix: Full Code Implementation}
	
	To ensure transparency and reproducibility, the core Python implementation of our analysis is provided below.
	
	\subsection{Part 1: PCA and Data Analysis}
	\begin{lstlisting}[language=Python, caption={PCA Variance Analysis Code}]
		import pandas as pd
		from sklearn.decomposition import PCA
		from sklearn.preprocessing import StandardScaler
		import matplotlib.pyplot as plt
		
		# Data Loading
		data_numeric = data_cleaned.drop(columns=['Entity'])
		scaler = StandardScaler()
		data_scaled = scaler.fit_transform(data_numeric)
		
		# PCA Execution
		pca = PCA()
		pca.fit(data_scaled)
		
		# Variance Analysis
		for i, (var, cum) in enumerate(zip(pca.explained_variance_ratio_, pca.explained_variance_ratio_.cumsum())):
		print(f"PC {i+1}: Variance={var:.4f}, Cumulative={cum:.4f}")
	\end{lstlisting}
	
	\subsection{Part 2: Dynamic Simulation (RK4 Implementation)}
	\begin{lstlisting}[language=Python, caption={Runge-Kutta 4 Simulation Algorithm}]
		import numpy as np
		import matplotlib.pyplot as plt
		
		# Saturation Function (prevents unbounded growth)
		def sigma(x):
		return x / (1 + x**2)
		
		# Differential Equation (Extended Lotka-Volterra)
		def dx_dt(x, r, A):
		interaction = np.dot(A, sigma(x))
		# Linear intervention term g(x) = 0.1x
		return x * (r + interaction) + 0.1 * x
		
		# RK4 Numerical Solver
		def rk4_step(x, r, A, dt):
		k1 = dx_dt(x, r, A)
		k2 = dx_dt(x + 0.5 * dt * k1, r, A)
		k3 = dx_dt(x + 0.5 * dt * k2, r, A)
		k4 = dx_dt(x + dt * k3, r, A)
		return x + (dt / 6.0) * (k1 + 2*k2 + 2*k3 + k4)
		
		# Simulation Loop (Conceptual)
		# for t in range(steps):
		#     x = rk4_step(x, r, A, dt)
	\end{lstlisting}
	
	\end{document}